\documentclass{article}

\usepackage[utf8x]{inputenc}
\usepackage{amsmath,amssymb,latexsym,amsfonts}
\usepackage{amsthm}
\usepackage[only,llbracket,rrbracket]{stmaryrd}
\usepackage{extarrows}
\usepackage{ifthen}
\usepackage[textsize=small]{todonotes}
\usepackage{paralist}
\usepackage{xifthen}
\usepackage{subfig}
\usepackage{colonequals}
\usepackage{mathabx}

\usepackage{hyperref}

\hypersetup{
    bookmarks=true,         
    unicode=false,          
    pdftoolbar=true,        
    pdfmenubar=true,        
    pdffitwindow=false,     
    pdfstartview={FitH},    
    pdftitle={My title},    
    pdfauthor={Author},     
    pdfsubject={Subject},   
    pdfcreator={Creator},   
    pdfproducer={Producer}, 
    pdfkeywords={keyword1} {key2} {key3}, 
    pdfnewwindow=true,      
    colorlinks=true,        
    linkcolor=black,        
    citecolor=black,        
    filecolor=magenta,      
    urlcolor=cyan           
}

\usepackage{thmtools,thm-restate}

\newcommand{\ignore}[1]{}
\newcommand{\ATA}{\mathcal}
\newcommand{\prestar}[1]{\mathrm{Pre}^*({#1})}

\usepackage{tikz}
\usetikzlibrary{arrows,shapes,snakes,automata,backgrounds,fit,positioning}

\tikzset{
  treenode/.style = {align=center, inner sep=0pt, text centered,
    font=\sffamily},
  arn_n/.style = {treenode, circle, white, font=\sffamily\bfseries, draw=black,
    fill=black, text width=1.5em},
  arn_r/.style = {treenode, circle, red, draw=red, 
    text width=1.5em, very thick},
  arn_x/.style = {treenode, rectangle, draw=black,
    minimum width=0.5em, minimum height=0.5em}
}

\usepackage{environ}
\NewEnviron{killcontents}{}

\title{Unified Analysis of Collapsible and Ordered Pushdown Automata via Term Rewriting}

\author{Lorenzo Clemente}

\begin{document}

\maketitle

\begin{abstract}
	
	We model collapsible and ordered pushdown systems with term rewriting,
	by encoding higher-order stacks and multiple stacks into trees.
	We show a uniform inverse preservation of recognizability result for the resulting class of term rewriting systems,
	which is obtained by extending the classic saturation-based approach.
	This result subsumes and unifies similar analyses on collapsible and ordered pushdown systems.
	Despite the rich literature on inverse preservation of recognizability for term rewrite systems,
	our result does not seem to follow from any previous study.

\end{abstract}

\paragraph{Introduction.}

Modelling complex systems requires to strike the right balance between the accuracy of the model,
and the complexity of its analysis.
A particularly successful case is that of \emph{pushdown systems} \cite{BouajjaniEsparzaMaler:Pushdown:1997,FinkelWillemsWolper:Pushdown:1997},
which can be used to model programs with recursive procedures.
A pushdown is an unbounded data structure which makes pushdown systems more expressive than finite-state systems,
and yet their reachability analysis still admits efficient (i.e., polynomial time) algorithms.
A standard technique to solve the reachability problem for pushdown systems
is the so-called \emph{saturation-based approach}
\cite{BouajjaniEsparzaMaler:Pushdown:1997,FinkelWillemsWolper:Pushdown:1997},
where ordinary finite automata representing potentially infinite sets of pushdown configurations
are manipulated until the full reachability set has been computed.

There has recently been an effort to model more complex features of recursive programs.
In one line of research, \emph{multi-threaded recursive programs},
i.e., concurrent programs where each thread consists of a recursive program,
have been modelled by multi-pushdown systems.
Since two pushdowns can already simulate the behaviour of a Turing machine (cf. \cite{Ramalingam:Undecidable:2000}),
various restrictions have been proposed to obtain a model with a decidable reachability analysis.
Instances include \emph{context-bounding} \cite{QadeerRehof:Context-Bounded:2005},
where the automaton can access a different stack only a bounded number of times,
the more general \emph{phase-bounding} \cite{LaTorreMadhusudanParlato:Bounded-phase:2007},
with a similar restriction but only on pop operations (i.e., push operations are always allowed),
and \emph{scope-bounding} \cite{latorre:napoli:CONCUR2011},
which generalises context-bounding in a different direction by requiring a bounded number of contexts only within the same scope
(i.e., between a push and its matching pop).
Yet another restriction is to consider \emph{ordered multi-pushdown systems}
\cite{BreveglieriCherubiniCitriniCrespi-Reghizzi:Ordered:1996} (later corrected in \cite{AtigBolligHabermehl:Ordered:2008}),
where the pushdowns are linearly ordered, push operations are unrestricted,
but a pop on a stack destroys the contents of all previous stacks.
Reachability analysis of ordered multi-pushdown systems is 2EXPTIME-complete \cite{AtigBolligHabermehl:Ordered:2008},
and an optimal saturation-based procedure has been proposed by Atig \cite{Atig:ordered:2012}.

In another line of research, \emph{higher-order recursive programs} are addressed,
i.e., programs where functions can be passed as parameters to other functions.
Higher-order features like this are nowadays present in all major programming languages,
such as C++, Java, Haskell, Python, Scala, Scheme, Erlang, and others.
In order to model higher-order recursion,
pushdown systems have been generalised to \emph{collapsible pushdown systems} \cite{HagueMurawskiOngSerre:Collapsible:2008}
(later simplified to \emph{annotated pushdown systems} \cite{BroadbentCarayolHagueSerre:Saturation:2012}),
where the pushdown is now a nested stack-of-stacks structure,
and basic symbols carry links/annotations to the state of the pushdown at the time when they where pushed on the stack.
Reachability analysis of order-$n$ collapsible/annotated pushdown systems is $(n-1)$-EXPTIME-complete,
where ordinary pushdown systems correspond to the order-$1$ case.
Also in this case, an optimal saturation-based procedure has been proposed by Broadbent et al. \cite{BroadbentCarayolHagueSerre:Saturation:2012},
and later implemented in the tool C-SHORe \cite{BroadbentCarayolHagueSerre:C-SHORe:2013}.

\paragraph{Inverse preservation of recognisability.}

The two decidability results above by Atig and Broadbent et al. for, resp., ordered and collapsible/annotated pushdown systems
are instances of what is called \emph{inverse preservation of recognisability} in the area of term-rewrite systems.
More precisely, by representing a possibly infinite set of target configurations $T$ by some extended finite automaton $\ATA A$,
the saturation procedure produces an automaton $\ATA B$ of the same kind recognising the backward reachability set $\prestar T$,
i.e., all those configurations that can reach $T$ in a finite number of steps.
This approach requires to either encode generalised pushdowns into ordinary strings and use standard finite automata,
or to develop ad-hoc extensions of finite automata working directly on generalised pushdowns.

The two approaches are equivalent, but have different merits.
Encoding multi-/annotated pushdowns into ordinary strings has the advantage to use a standard notion of recognisability of sets of configurations
at the expense of a cumbersome saturation procedure.
Seth gives a simple encoding of multi-pushdowns into simple strings \cite{Seth:Global:2010}.
The result by Atig \cite{Atig:ordered:2012} shows inverse preservation of recognisability for sets of configurations of ordered pushdown automata,
where configurations are linearised with the encoding of Seth.

\ignore{
For example, in an ordered pushdown system, a configuration is a tuple $(p, w_1, \dots, w_n)$,
where $p$ is a control state and $w_1, \dots, w_n$ are the contents of pushdowns $1, \dots, n$, respectively.
One approach to represent sets of configurations as above by finite automata
is to encode the contents of the pushdowns as the word $w_1 \$ \cdots \$ w_n$,
for a new separator symbol $\$$ \cite{Seth:Global:2010}.
With this encoding, one can define the notion of recognisable set of configurations by using standard finite automata,
and \cite{Atig:ordered:2012} basically proves an inverse preservation of recognisability result for this encoding.
}

On the other side, using a extended notion of finite automaton specialised to work directly on generalised pushdowns
has the advantage that the saturation procedure has a much more natural presentation,
at the expense of the need to define a new, ad-hoc notion of finite automata.
This is the approach followed by Broadbent et al. \cite{BroadbentCarayolHagueSerre:Saturation:2012},
who define a notion of higher-order stack automata to read in a nested fashion annotated pushdowns.
Note that higher-order stack automata can be converted into ordinary finite automata
by using the string encoding of annotated pushdowns proposed by \cite{BouajjaniMeyer:Higher:2007}.
Therefore, higher-order stack automata still recognise only recognisable sets of annotated pushdown configurations.
The result by Broadbent et al. \cite{BroadbentCarayolHagueSerre:Saturation:2012}
shows inverse preservation of recognisability for sets of configurations of annotated pushdown automata.

\ignore{
The encoding for annotated pushdown systems is more complicated.
An order-$0$ pushdown is just a basic symbol $a$,
and, for $n > 0$, an order-$n$ pushdown is a sequence $w_1, \dots, w_k$ of order-$(n-1)$ pushdowns.
By introducing bracket symbols ``$[$'' and ``$]$'', the sequence $w_1, \dots, w_k$ is encoded as the bracketed word $[w'_1 \dots w'_k]$,
where the $w'_j$'s are encoded recursively.
In this way, recognisable sets of annotated pushdowns can be defined with standard finite automata,
and \cite{BroadbentCarayolHagueSerre:Saturation:2012}
proves an inverse preservation of recognisability result for this encoding.
}

\paragraph{A term-rewriting approach.}

We propose to use tree automata techniques to provide a solution
covering both the results of Atig and Broadbent et al.
in a uniform framework.
We propose to encode multi- and annotated pushdowns as \emph{trees} (i.e., first-order terms),
and to use ordinary tree automata to represent sets of configurations.
Then, transition rules for multi- and annotated pushdown systems can be given directly on these trees
in the form of \emph{root-rewriting rules} of a restricted class.
Finally, by instantiating a generic saturation procedure on the resulting rewrite rules
we can derive in an uniform way the inverse preservation of recognisability results of Atig and Broadbent et al..

This has the following advantages:
\begin{inparaenum}
	\item We do not need to introduce ad-hoc finite automata models to recognise generalised pushdown configurations.
	We use standard finite tree automata, which comes with the standard notion of recognisable set of finite trees.
	\item We obtain simple saturation rules, thanks to the simplicity of the encoding into trees.
	\item Our procedure has optimal complexity when instantiated to the specific instances provided by ordered and annotated pushdown systems.
\end{inparaenum}

\paragraph{Discussion.}

Our inverse preservation of recognisability does not seem to follow from the rich literature on term-rewriting systems.
Indeed, most results of this kind were proved for different kinds of \emph{bottom-up rewriting}:
Starting from the result of Brainerd on \emph{ground rewrite systems} \cite{Brainerd:Ground:1969},
increasingly more general classes of term-rewrite systems were proved to preserve or inverse preserve recognisability,
such as \emph{monadic rewrite systems} \cite{Salomaa:Monadic:1988},
\emph{semi-monadic rewrite systems} \cite{CoquideDauchetGilleronVagvolgyi:Semi-monadic:1994},
\emph{shallow rewrite systems} \cite{Comon:Shallow:2000},
\emph{growing rewrite systems} \cite{Jacquemard:Growing:1996,NagayaToyama:Left-LinearGrowing:2002},
\emph{generalised semi-monadic rewrite systems} \cite{GyenizseVagvolgyi:GeneralisedSemi-Monadic:1998},
\emph{finite-path overlapping rewrite systems} \cite{TakaiKajiSeki:FinitePathOverlapping:2000},
culminating in the most general class of \emph{bottom-up rewrite systems} \cite{SenizerguesDurand:BottomUp:2007,DurandSylvestre:LeftLinearBottomUp:2011}.

Instead, we use root-rewriting, and our term-rewrite systems can be seen as a generalisation on trees of \emph{prefix-rewrite systems}.
Preservation of recognisability results for prefix-rewrite systems on words
were known since B\"uchi \cite{Buchi:Canonical:1964}
(cf. also \cite{Benois:Preservation:1987,Caucal:PrefixRewriting:1992} and the book by Book \& Otto \cite{BookOtto:StringRewriting:1993}).

Finally, a class of root-rewriting systems called \emph{generalised growing rewrite systems} has been proposed \cite{NittaSeki:GeneralisedGrowing:2003}.
Under some additional condition, this class is shown to enjoy inverse preservation of recognisability.
Generalised growing rewrite systems are closer in spirit to our class than others,
since both use root-rewriting.
However, the two classes turn out to be syntactically incomparable.

\paragraph{Acknowledgements.}

We kindly acknowledge many inspiring discussions on this topic with Igor Walukiewicz.

\bibliography{../../../../mybibdatabase}

\end{document}